# Heuristics guide the implementation of social preferences in one-shot Prisoner's Dilemma experiments

Valerio Capraro[1], Jillian J. Jordan[2], David G. Rand[2,3,4*]

First version: April 28, 2014

This version: August 12, 2014

[1]Department of Mathematics, University of Southampton, Southampton, UK, [2]Department of Psychology, [3]Department of Economics, [4]School of Management, Yale University, New Haven CT 06511 USA

*Corresponding author: David.Rand@Yale.edu

Cooperation in one-shot anonymous interactions is a widely documented aspect of human behaviour. Here we shed light on the motivations behind this behaviour by experimentally exploring cooperation in a one-shot continuous-strategy Prisoner's Dilemma (i.e. one-shot two-player Public Goods Game). We examine the distribution of cooperation amounts, and how that distribution varies based on the benefit-to-cost ratio of cooperation (b/c). Interestingly, we find a trimodal distribution at all b/c values investigated. Increasing b/c decreases the fraction of participants engaging in zero cooperation and increases the fraction engaging in maximal cooperation, suggesting a role for efficiency concerns. However, a substantial fraction of participants consistently engage in 50% cooperation regardless of b/c. The presence of these persistent 50% cooperators is surprising, and not easily explained by standard models of social preferences. We present evidence that this behaviour is a result of social preferences guided by simple decision heuristics, rather than the rational examination of payoffs assumed by most social preference models. We also find a strong correlation between play in the Prisoner's Dilemma and in a subsequent Dictator Game, confirming previous findings suggesting a common prosocial motivation underlying altruism and cooperation.

Cooperation is central to human societies, from personal relationships to workplace collaborations, from environmental conservation to political participation, international relations, and price competition in markets[1-17]. A simple model commonly used to study cooperation is the Prisoner's dilemma (PD), in which two agents can either cooperate (C) or defect (D): cooperating means paying a cost $c$ to give a benefit $b$ ($b>c$) to the other person; defecting means doing nothing. The PD is an attractive model of cooperation because it highlights the tension between individual and collective interests: agents maximize their personal payoff by defecting (and avoiding the cost of cooperation). But if both agents defect, both are worse off than if they had both cooperated.

Since cooperation is individually costly, standard economic models predict that people should not cooperate (unless the game is repeated, in which case theoretical models predict[18-22], and behavioural experiments demonstrate[23-29], that cooperation can be favoured via 'reciprocity'; in repeated games, even selfish players may cooperate in order to gain the benefits of reciprocal cooperation in future periods[30]). Yet cooperation in one-time encounters with strangers is common outside the laboratory, and a substantial amount of cooperative behaviour is observed in one-shot PD experiments in the lab with anonymous players[31-41].

Here we attempt to go beyond the observation that people sometimes cooperate in one-shot anonymous PDs by shedding new light on the *motivations* underlying this cooperative behaviour. Rather than giving participants a binary choice between C or D as is typically done, we make the decision space continuous (i.e. use a continuous-strategy PD): each participant chooses how much of an endowment to spend on helping the other player, with every $c$ units spent resulting in the other person gaining $b$ units. We also vary the $b/c$ ratio, and ask how the distribution of cooperation levels changes as a result. This allows us to evaluate the predictions of different theories of cooperative behaviour and gain insight into the underpinnings of cooperation in the one-shot PD.

The standard explanation in economics for non-zero cooperation in one-shot games involves *social preferences*. Social preference theories typically assume that people are rational, but that their utility functions include more than just their own material payoff. Three main types of social preferences have been proposed: efficiency[42], whereby people get utility from aggregate welfare (i.e. total payoff of all players) and thus may be willing to pay costs to give large benefits to others; inequity aversion[43,44], whereby people get disutility from unequal payoffs and thus may be willing to pay to reduce the difference between their payoff and the payoffs of others; and reciprocity[45], whereby people get utility from cooperating with those who are cooperative and not cooperating with (or punishing) those who are uncooperative, and thus may be willing to pay the cost of cooperation if they expect others to do the same.

Efficiency models make a clear prediction regarding the distribution and *b/c* dependence of cooperation levels in a continuous-strategy PD: players who primarily care about efficiency should engage in zero cooperation if *b/c* is below the critical threshold at which it becomes worth it for them to cooperate, and should engage in maximal cooperation if *b/c* is above this threshold. As threshold values vary across participants, increasing *b/c* should increase the average level of cooperation by shifting more participants from zero cooperation to maximal cooperation.

Theories based on inequity aversion and reciprocity, conversely, do not make clear predictions, either about the distribution of cooperation levels or about the response to changes in *b/c*. Both inequity aversion and reciprocity favour matching the cooperation level of one's partner; thus any level of cooperation could be supported by these preferences, depending on one's expectations (i.e. 'beliefs') about the behaviour of the partner. (This includes zero cooperation: if I believe my partner is self-interested, I will not cooperate even though I have a social preference for equality or reciprocity.) Therefore, as people can be expected to differ in their expectations about the cooperation levels of their partners (given variance in past experience inside and outside of the laboratory setting), these models predict a range of different cooperation levels, with no reason to expect specific levels to be more common than others. Furthermore, inequity averse and reciprocal participants will only change their cooperation level in response to changes in *b/c* in so much as they expect *b/c* to change the behaviour of their partner (for example, if they assume their partner has some preference for efficiency). Thus an increase in cooperation with *b/c* is an indication of participants either having efficiency preferences, or expecting others to have such preferences.

We also evaluate predictions generated by another class of models which relax the rationality assumption of standard social preference models. There is considerable evidence that heuristics, rather than rational utility maximization, play an important role in decision-making[46-50]. Heuristics are simple rules of thumb prescribing behaviour which is typically desirable, but is not precisely tuned to the details of the current decision. Such heuristics may interact with social preferences in various ways. For example, inequity averse people might seek equal outcomes based on a fairness heuristic which favours equal splits of the endowment, even in cases where an equal split does not actually lead to equal payoffs (e.g. if money transferred to the other person is multiplied by a constant, *b/c*>1, transferring half of the endowment causes the other person to earn more). In addition, people with inequity averse or reciprocal preferences who are trying to predict the behaviour of their partner might be influenced by a heuristic that leads them to settle on particularly salient values such as the mid-point of the scale (50% cooperation) rather than carefully reasoning about what partner behaviour is most likely given the *b/c* ratio. Both of these interactions between heuristic reasoning and social preferences predict relative insensitivity to *b/c*, as well as a distribution of cooperation levels with substantial weight concentrated at 50%.

In sum, we therefore expect some participants to cooperate either fully or not at all, and for increasing *b/c* to increase the proportion of these participants choosing full cooperation; whereas we expect other participants to engage in 50% cooperation and to be insensitive to *b/c*.

To evaluate these various predictions, we had 308 participants play a continuous-strategy PD in which they were given ten monetary units, and then decided how many to transfer to their partner, with any transferred units being multiplied by a constant (the *b/c* value). We varied the *b/c* ratio across *b/c*=[2,3,4,5,10], with each participant making only a single decision with a single *b/c*. Finally, we sought to replicate recent results regarding the 'cooperative phenotype'[51], which suggest that a common motivation underlies both cooperation in the PD and altruism. Thus, after they completed the PD, we had participants play a unilateral, zero-sum money transfer (i.e. Dictator Game, DG).

**Results**

The distribution of cooperation levels for each *b/c* value is shown in Figure 1. For all values of *b/c*, we see a strongly tri-modal distribution concentrated on 'give nothing', 'give half', and 'give everything'. Aggregating over all *b/c* values, we find that 22.4% participants transferred nothing, 19.2% participants transferred half, 52.3% participants transferred all, and only 6.2% participants transferred other amounts.

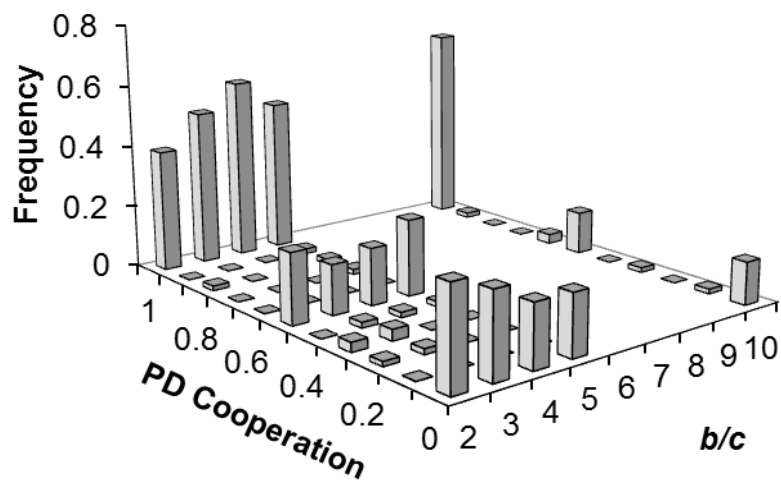

*Figure 1. Distribution of cooperation levels in the PD as a function of benefit-to-cost ratio.*

We next ask how the probabilities of giving nothing, half, and everything change with *b/c* using logistic regression. We find (i) that participants are significantly less likely to give nothing as *b/c* increases (coeff = -.130, p = .019); (ii) that participants are significantly more likely to give everything as *b/c* increases (coeff = .106, p = .010); and (iii) that the probability of giving half does not change with *b/c* (coeff = -.056, p = .302). These results are robust to controlling for age, gender, education, and log-transformed number of previous studies completed (Probability of giving nothing: coeff = -.145, p = .013; giving everything: coeff = .131, p = .003; giving half: coeff = -.085, p = .142); for completeness we report that when including demographics we also find that women (coeff = -0.779, p=0.003) and participants who have had more experience with economic games (coeff = -0.330, p = 0.030) are significantly less likely to transfer everything. Furthermore, we do not find evidence of diminishing returns on increasing *b/c*: when redoing all of the above regressions including a $(b/c)^2$ term (to capture non-linear effects of *b/c*), the non-linear term is never significantly different from zero, p>0.3 for all. Thus it appears that increasing *b/c* shifts people from transferring nothing to transferring everything, without affecting the percentage who give an even split of half.

We now examine how the mean level of cooperation changes as a function of *b/c* (Figure 2). Linear regression finds a significant positive relationship between cooperation and *b/c* (coeff = .239, p = .003; including controls: coeff = .258, p = .002); for completeness we report that when including demographics we find similar results as above, with women (coeff = -1.12, p = 0.022) and participants have more experience with economic games (coeff = -0.592, p = 0.041) being less cooperative on average (these findings related to experience are consistent with previous work showing that experience with economic games undermines cooperative intuitions[50,52]). We again find no evidence of a non-linear relationship between *b/c* and cooperation (including $(b/c)^2$ term, p = 0.202). We also note that the relationship between mean cooperation and *b/c* is robust to excluding participants who made transfers other than nothing or everything (coeff = .283, p = .007).

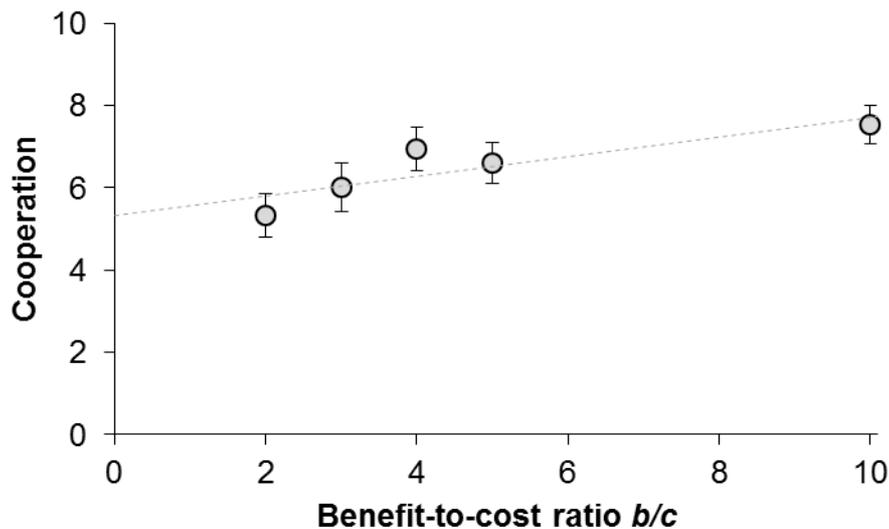

*Figure 2. Average amount of cooperation in the PD as a function of benefit-to-cost ratio. Error bars indicate standard errors of the mean.*

Finally, we analyse the relationship between cooperation in the PD and giving in the subsequent DG (Figure 3). Aggregating across conditions, we find a strong positive association between the PD and the DG (pairwise correlation: r=.561, p<.001; linear regression predicting DG as a function of PD, coeff=.522, p<.001; with controls: coeff=.535, p<.001; no significant correlation between any of the controls and DG giving). Examining Figure 3 shows that this correlation is largely driven by a lack of participants who gave in the DG but did not cooperate in the PD. Put differently, cooperators were not necessarily DG givers, but DG givers were almost certainly PD cooperators.

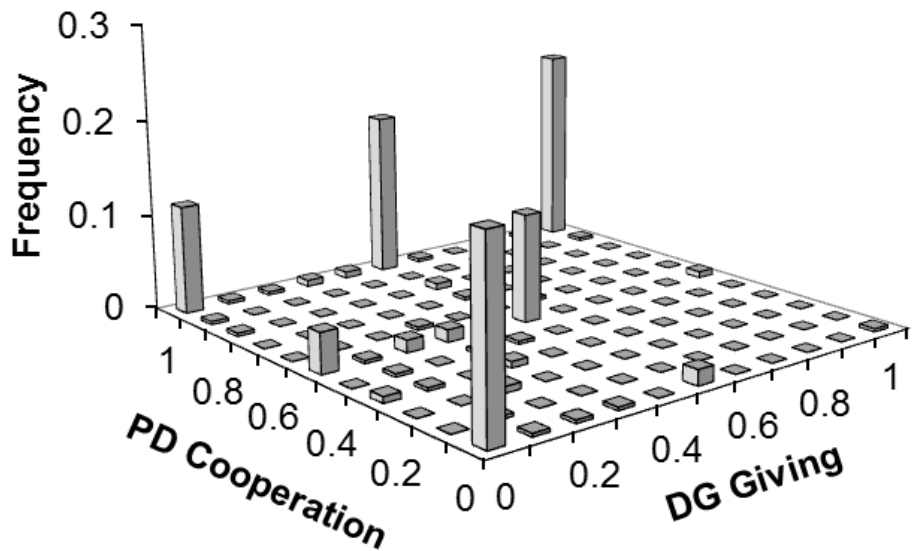

*Figure 3. Joint distribution of cooperation in the PD and giving in the DG.*

**Discussion**

Here we have shown that cooperation levels in a one-shot continuous-strategy PD are tri-modally distributed, with peaks at zero, half and full cooperation; and that increasing the *b/c* ratio reduces zero cooperation and increases full cooperation, but that the influence of the *b/c* ratio is somewhat limited. Further, we have shown that giving in the PD is strongly correlated with giving in a subsequent DG.

The trimodal distribution we observe is not readily consistent with predictions of standard social preference models in which participants rationally maximize utility functions that depend on the payoffs of others. Efficiency concerns clearly predict a bimodal pattern of zero or full cooperation, with *b/c* decreasing zero and increasing full cooperation. Seeing as no information was given about the distribution of cooperation levels, inequity averse and reciprocal players would presumably have a range of different beliefs regarding their partner's expected behaviour. As a result, these models would predict a wide range of different cooperation levels, with no reason to expect clear modes at 0%, 50% or 100%. Furthermore, if these players believe that *other* players may have efficiency preferences, then they should anticipate partner cooperation increasing with *b/c* and therefore increase their own cooperation levels accordingly. Thus the modes at 0% and 100% cooperation, and the increase in cooperation with *b/c*, suggest that some participants either have efficiency preferences or anticipate that others will have efficiency preferences.

The observed *tri*modal distribution with substantial weight at 50%, and the relatively small increase in cooperation in response to a large increase in *b/c*, conversely, are surprising in light of traditional social preference models. Both of these features, however, are direct predictions of theories based on heuristic reasoning. A simple fairness heuristic could lead participants to transfer 50% to their partners, mistakenly believing that this would lead to equal payoffs. (Note that in the PD, unlike in the Dictator Game, 50% cooperation is not the naturally equitable choice: this is both because there is multiplier on transfers in the PD, such that if you give 5 out of 10 units, the other person receives 5*b/c units; and because the other person is also making a decision.) Or a heuristic could lead expectations regarding the partner's behaviour to naively anchor on the salient mid-point of 50%.

To gain greater insight into the motivation of participants engaging in 50% cooperation, we examined responses these participants gave at the end of the study to the prompt "please describe why you made your decision in the game" (such free-response texts can give useful insights into participants' decision processes in economic games[53,54]). Consistent with a fairness heuristic, 24% of statements explicitly mentioned a desire to be fair as the main motivator for their choice to transfer half their endowment to the partner. Consistent with a heuristic focusing beliefs regarding partner behaviour on the salient scale midpoint of 50% cooperation, 15% of statements explicitly said they engaged in 50% cooperation because they expected their partner to engage in 50% cooperation, while an additional 22% of statements said that they were unsure of whether their partner would cooperate and therefore only transferred half of the endowment. (Interestingly, another 16% of statements explained the choice to transfer half of the endowment by saying that it was a compromise between generosity and self-interest, a motivation that to our knowledge has not been previously discussed and which merits further study; the remaining 24% of statements either gave no reason or were not readily categorizable). In sum, participants' post-experimental descriptions of their decision processes provide direct evidence of heuristic use motivating the choice to engage in 50% cooperation. We note that many of the participants engaging in zero or full cooperation may have also been using heuristics, but for these choices it is difficult to disentangle heuristic reasoning from rational application of social preferences, as they lead to the same outcomes. For example, a heuristic that prescribes contributing everything is indistinguishable in this paradigm from a rationally applied efficiency preference. (This is unlike a fairness heuristic that prescribes giving half of the endowment, because giving half does not in general actually create equal outcomes in our PDs.) Finally, it is important to note that the present study illuminates the role heuristics play in the *implementation* of social preferences, rather than the role that heuristics formed via internalization of norms may play in the *origin* of social preferences[50,55,56].

An important limitation of our experiment is that because of our between-subjects design, we cannot observe *specific* individuals changing their behaviour. Thus we cannot distinguish between two possibilities among our participants giving nothing or everything: it could be that each such person has a personal minimum *b/c* at which the psychological benefits of cooperation begin to outweigh the financial costs. If this was the case, any given person would always give nothing below that critical *b/c*, and always give everything above it; and the gradual increase in average cooperation with *b/c* we observe in Figure 2 would be the result of more and more people having passed their personal thresholds. Alternatively, it could be that people behave probabilistically, with their chance of cooperating in any given decision increasing as *b/c* increases. In this case, the graded response to *b/c* that we observe at the population level in Figure 2 would also be reproduced within each individual. Distinguishing between these possibilities is likely to be difficult, however, because of consistency and contagion effects, like those we observed between the PD and the DG: if one's choice in a given cooperation decision is heavily influenced by choices in immediately previous decisions, it makes it difficult for experimenters to obtain a clean measure of how the payoff structure influences that person's choices. Nonetheless, this is an important direction for future research.

To our knowledge, only a handful of previous studies have experimentally investigated how the payoff structure affects cooperation in a one-shot PD. All of these studies have used a binary PD, preventing them from drawing conclusions regarding the distribution of cooperation levels, which is our main focus. With respect to the effect of the PD's payoff structure on cooperation, these studies have typically not used the benefit-to-cost ratio decomposition of PD payoffs, but instead directly varied one or more of the payoffs associated with the four possible PD outcomes ([C,C],[C,D],[D,C],[D,D]). An early study found that cooperation increased as the [D,D] payoff was decreased[57], a finding that was replicated across a wider range of values in a more recent study[34]. Two other studies found that cooperation decreased as the ratio of payoffs ([D,C]-[D,C])/([C,C]-[D,D]) was increased[58,59]. We add to these studies by examining the distribution of cooperation amounts, and by using the *b/c* formulation which is standard in evolutionary game theory[60] and readily interpretable in terms of predictions based on efficiency preferences.

The substantial correlation we observe between play in the PD and the DG adds weight to previous work from our group showing significant correlations in play across the DG, the Public Goods Game (a 4-person version of our continuous PD), and the Trust Game, which was argued to reflect a 'cooperative phenotype'[51]. This replication is important, given that an earlier study found no correlation between the Public Goods Game and a modified Dictator Game in which participants made 21 decisions between two pairs of options that were more or less fair[61]. Both our study and ref 51 had an order of magnitude more participants than ref 61; thus, it is possible that the latter null result was due to a lack of power. It could also be that the modified DG structure of ref 61, in which many extremely similar decisions were made in a row, introduced self-consistency effects or other confounds that obscured a true relationship with the Public Goods Game.

Further evidence regarding the relationship between cooperation and fairness comes from the fact that DG givers in our study were almost entirely a strict subset of PD cooperators. This observation suggests that the motives present in the DG (e.g. inequity aversion) are also present in the PD, but that additional motives exist in the PD that do not in the DG (e.g. concerns about efficiency or the choice of the other player).

In sum, our results give insight into the decision-making process in one-shot anonymous Prisoner's Dilemma games. We provide evidence that many people who cooperate deviate from traditional models of rational self-interest not only by being sensitive to the payoffs of others (i.e. being 'other-regarding'), but also by using simple heuristics. We also provide further evidence for a domain general proclivity to cooperate across games.

**Methods**

We recruited participants using the online labour market Amazon Mechanical Turk (MTurk)[32,62-64]. Participants received a $0.35 show-up fee and were told they would be playing a two-stage game in which they could earn additional income.

In the first stage, participants were paired with another MTurk worker, and both were given $0.10. They each then chose how much, if any, to transfer to the other person, with any transfers being multiplied by a constant $k$. We manipulated the PD payoff structure by varying the value of $k$ across $k=[2,3,4,5,10]$, with a given participant being randomly assigned to a single value of $k$ (i.e. a between-subjects design). In this continuous PD, $b/c=k$ because for each cent participants transferred, the recipient received $k$ cents.

Before making their decision, participants answered comprehension questions to make sure they understood the payoff structure (see Appendix for the exact instructions). Given that our key manipulation involved changing the payoff structure, it was essential that participants understood the payoffs. Therefore, participants who answered any questions incorrectly were not allowed to participate. After answering the comprehension questions, participants made their PD decision, and then moved on to the second stage (without learning their partner's decision in the PD, to prevent contagion effects).

In the second stage, participants were paired with a different MTurk worker. Participants were given $0.10 and had to decide how much, if any, to unilaterally transfer to the other person (transfers were not multiplied, and the other person had no initial endowment and made no transfer decision – i.e. participants played a standard Dictator Game). Finally, participants completed a free-response describing the reasons for their decisions in the games and a demographic survey. After all participants had been recruited, they were matched at random and payoffs were calculated as described. No deception was used in this study, informed consent was obtained from all participants, and the study was approved by the Harvard University Committee on the Use of Human Subjects. Methods were carried out in accordance with the approved guidelines.

A total of 308 US resident participants answered all comprehension questions correctly (mean age=30.8 years, 62% male) and were thus allowed to participate in the experiment (140 people answered one or more comprehension questions incorrectly and were excluded). IP addresses were screened to prevent the same person from participating repeatedly. 66 participants were assigned to the PD with multiplier k=2; 56 participants were assigned to the PD with multiplier k=3; 60 participants were assigned to the PD with multiplier k=4; 61 participants were assigned to the PD with multiplier k=5; and 65 participants were assigned to the PD with multiplier k=10. In addition to the $0.35 show up fee for completing the task, participants earned an average of $0.47 of additional income based on the games.

To assess the free-response statements participants giving half in the PD provided regarding their motivations in the game, two research assistants coded each response from these participants. The coders were not informed about the purpose of the study or the various hypothesis and predictions being tested. For each statement, they were asked which of the following eight categories best described it (fraction of statements assigned to each category indicated in parentheses):

1. The participant explicitly said that they took the action because it was fair (24%)
2. The response indicates that the participant expected the other person to take the same action, so that's why they took it (15%)
3. The response indicates that because the participant felt uncertainty about the other person's action, the participant decided to hedge / reduce variance by taking this action (22%)
4. The response indicates that the participant wanted to compromise between taking a selfish action and taking a generous action (16%)
5. The participant explicitly references intuition / their gut feeling / going with the first thing that came to them (3%)
6. The response indicates that the group payoff would be maximized by taking this action / it would be overall best for everyone to take this action (2%)
7. The response restates what the person did, but does not provide an explanation (13%)
8. The response indicates that the participant didn't want to be greedy / selfish (6%)

Finally, we show the instructions of the $k=10$ condition here:

> Welcome to this HIT. This HIT has two parts. We will tell you about the second part after you have completed the first one.
>
> For your participation in this HIT, you receive 30 cents. You also can earn a bonus as described on the following pages.
>
> How much of a bonus depends on your own decisions and also on the decision of an anonymous other MTurk participant with whom you are paired.
>
> You will be told about the outcome of all parts of the HIT at the time your bonus is paid.

> This is the first part of the HIT.
>
> You are together with another, anonymous participant. How much money you earn depends on your own choice, and on the choice of the other participant.
>
> You are given 10 additional cents and you have to decide how much, if any, to transfer to the other participant. The amount of money you decide to transfer will be multiplied by 10 and earned by the other participant.
>
> The other participant will be given the same choice.
>
> So, for instance, if you both transfer everything, you both get 1 dollar and are better off than if you both keep everything. If you transfer everything and the other person keeps everything, then you earn nothing. If you keep everything and the other participant transfers everything, then you get 1 dollar and 10 cents.
>
> The other person is REAL and will really make a decision. Once you have each made your decision, neither of you will ever be able to affect each others' bonuses in later parts of the HIT.
>
> Now we will ask you several questions to make sure that you understand how the payoffs are determined.
> YOU MUST ANSWER ALL THESE QUESTIONS CORRECTLY TO RECEIVE A BONUS!
>
> Which action by YOU gives YOU a higher bonus?
> 0 cents  1 cent  2 cents  3 cents  4 cents  5 cents  6 cents  7 cents  8 cents  9 cents  10 cents
>
> Which action by YOU gives the OTHER PLAYER a higher bonus?
> 0 cents  1 cent  2 cents  3 cents  4 cents  5 cents  6 cents  7 cents  8 cents  9 cents  10 cents
>
> Which action by the OTHER PLAYER gives the OTHER PLAYER a higher payoff?
> 0 cents  1 cent  2 cents  3 cents  4 cents  5 cents  6 cents  7 cents  8 cents  9 cents  10 cents
>
> Which action by the OTHER PLAYER gives YOU a higher payoff
> 0 cents  1 cent  2 cents  3 cents  4 cents  5 cents  6 cents  7 cents  8 cents  9 cents  10 cents
>
> WHAT IS YOUR CHOICE?
> Tick the amount of money you want to transfer to the other participant.
> 0 cents  1 cent  2 cents  3 cents  4 cents  5 cents  6 cents  7 cents  8 cents  9 cents  10 cents

**This is the second part of the HIT.**

**You are paired together with another, anonymous participant, different from the one with whom you were paired in the first part of the HIT. This time how much money you earn depends only on your own choice.**

**You are given 10 additional cents and you have to decide how much, if any, to DONATE to the other participant.**

**The other participant has no choice: she or he will get your donation.**

**WHAT IS YOUR DONATION?**

**Tick the amount of money you would like to donate to the other participant.**

| 0 cents | 1 cent | 2 cents | 3 cents | 4 cents | 5 cents | 6 cents | 7 cents | 8 cents | 9 cents | 10 cents |
|---------|--------|---------|---------|---------|---------|---------|---------|---------|---------|----------|
| ○ | ○ | ○ | ○ | ○ | ○ | ○ | ○ | ○ | ○ | ○ |

**References**

1   Trivers, R. The evolution of reciprocal altruism. *The Quarterly Review of Biology* **46**, 35-57 (1971).
2   Axelrod, R. & Hamilton, W. D. The evolution of cooperation. *Science* **211**, 1390-1396 (1981).
3   Ostrom, E. *Governing the commons: The evolution of institutions for collective action*. (Cambridge Univ Pr, 1990).
4   Doebeli, M. & Hauert, C. Models of cooperation based on the Prisoner's Dilemma and the Snowdrift game. *Ecology Letters* **8**, 748-766 (2005).
5   Nowak, M. A. Five rules for the evolution of cooperation. *Science* **314**, 1560-1563 (2006).
6   Traulsen, A. & Nowak, M. A. Evolution of cooperation by multilevel selection. *Proc Natl Acad Sci USA* **103**, 10952-10955 (2006).
7   Hauert, C., Michor, F., Nowak, M. A. & Doebeli, M. Synergy and discounting of cooperation in social dilemmas. *Journal of Theoretical Biology* **239**, 195 (2006).
8   Crockett, M. J. The neurochemistry of fairness. *Annals of the New York Academy of Sciences* **1167**, 76-86 (2009).
9   Sigmund, K. *The calculus of selfishness*. (Princeton Univ Press, 2010).
10  Zaki, J. & Mitchell, J. P. Equitable decision making is associated with neural markers of intrinsic value. *Proceedings of the National Academy of Sciences* **108**, 19761-19766, doi:10.1073/pnas.1112324108 (2011).
11  Apicella, C. L., Marlowe, F. W., Fowler, J. H. & Christakis, N. A. Social networks and cooperation in hunter-gatherers. *Nature* **481**, 497-501 (2012).
12  Traulsen, A., Röhl, T. & Milinski, M. An economic experiment reveals that humans prefer pool punishment to maintain the commons. *Proceedings of the Royal Society B: Biological Sciences* **279**, 3716-3721 (2012).
13  Capraro, V. A Model of Human Cooperation in Social Dilemmas. *PLoS ONE* (2013).
14  Rand, D. G. & Nowak, M. A. Human cooperation. *Trends in Cognitive Science* **17**, 413-425 (2013).
15  Zaki, J. & Mitchell, J. P. Intuitive Prosociality. *Current Directions in Psychological Science* **22**, 466-470, doi:10.1177/0963721413492764 (2013).
16  Hauser, O. P., Rand, D. G., Peysakhovich, A. & Nowak, M. A. Cooperating with the future. *Nature* **511**, 220-223 (2014).
17  Jordan, J. J., Peysakhovich, A. & Rand, D. G. "Why we cooperate" in *The Moral Brain: Multidisciplinary Perspectives*   (eds J Decety & T Wheatley) (MIT Press, In press).
18  Nowak, M. A. & Sigmund, K. Tit for tat in heterogeneous populations. *Nature* **355**, 250-253 (1992).
19  Rand, D. G., Ohtsuki, H. & Nowak, M. A. Direct reciprocity with costly punishment: generous tit-for-tat prevails. *Journal of Theoretical Biology* **256**, 45-57 (2009).
20  Capraro, V., Venanzi, M., Polukarov, M. & Jennings, N. R. Cooperative equilibria in iterated social dilemmas. *Proceedings of the 6th International Symposium on Algorithmic Game Theory*, 146-158 (2013).
21  Nowak, M. A., Sasaki, A., Taylor, C. & Fudenberg, D. Emergence of cooperation and evolutionary stability in finite populations. *Nature* **428**, 646-650 (2004).

22  van Veelen, M., García, J., Rand, D. G. & Nowak, M. A. Direct reciprocity in structured populations. *Proceedings of the National Academy of Sciences* **109**, 9929-9934 (2012).

23  Dal Bo, P. & Frechette, G. R. Strategy choice in the infinitely repeated prisoners dilemma. *Working Paper*, SSRN 2292390 (2013).

24  Dal Bo, P. Cooperation under the shadow of the future: experimental evidence from infinitely repeated games *The American Economic Review* **95**, 1591-1604 (2005).

25  Dal Bo, P. & Frechette, G. R. The Evolution of Cooperation in Infinitely Repeated Games: Experimental Evidence *The American Economic Review* **101**, 411-429 (2011).

26  Blonski, M., Ockenfels, P. & Spagnolo, G. Equilibrium Selection in the Repeated Prisoner's Dilemma: Axiomatic Approach and Experimental Evidence. *American Economic Journal: Microeconomics* **3**, 164-192 (2011).

27  Dreber, A., Rand, D. G., Fudenberg, D. & Nowak, M. A. Winners don't punish. *Nature* **452**, 348-351 (2008).

28  Fudenberg, D., Rand, D. G. & Dreber, A. Slow to Anger and Fast to Forgive: Cooperation in an Uncertain World. *American Economic Review* **102**, 720-749 (2012).

29  Rand, D. G., Fudenberg, D. & Dreber, A. It's the thought that counts: The role of intentions in noisy repeated games. *Available at SSRN: http://ssrn.com/abstract=2259407* (2014).

30  Dreber, A., Fudenberg, D. & Rand, D. G. Who cooperates in repeated games? *Journal of Economic Behavior & Organization* **98**, 41-55 (2014).

31  Wong, R. Y. & Hong, Y. Y. Dynamic Influences of Culture on Cooperation in the Prisoner's Dilemma *Psychological Science* **16**, 429-434 (2005).

32  Horton, J. J., Rand, D. G. & Zeckhauser, R. J. The online laboratory: conducting experiments in a real labor market. *Experimental Economics* **14**, 399-425 (2011).

33  Artinger, F., Fleischhut, N., Levanti, V. & Stevens, J. R. Cooperation in risky environments: decisions from experience in a stochastic social dilemma. *Proceedings of the 34th Conference of the Cognitive Science Society*, 84-89 (2012).

34  Engel, C. & Zhurakhovska, L. When is the Risk of Cooperation Worth Taking? The Prisoners Dilemma as a Game of Multiple Motives. *Working Paper* (2012).

35  Engel, C. & Rand, D. G. What does "clean" really mean? The Implicit Framing of Decontextualized Experiments. *Economics Letters* (2013).

36  Khadjavi, M. & Lange, A. Prisoners and their dilemma *Journal of Economic Behavior & Organization* **92**, 163-175 (2013).

37  Rand, D. G. *et al.* Religious motivations for cooperation: an experimental investigation using explicit primes. *Religion, Brain & Behavior*, 1-18 (2013).

38  Camerer, C. *Behavioral Game Theory*.  (Princeton University Press, 2003).

39  Capraro, V., Smyth, C., Mylona, K. & Niblo, G. Benevolent characteristics promote cooperative behaviour among humans. *PLoS ONE* (In press).

40  Barcelo, H. & Capraro, V. Group size effect on cooperation in social dilemmas. *Available at SSRN: http://ssrn.com/abstract=2425030* (2014).

41  Capraro, V. & Marcelletti, A. Do good actions inspire good actions in others? *Available at SSRN: http://ssrn.com/abstract=2454667* (2014).

42  Charness, G. & Rabin, M. Understanding Social Preferences with Simple Tests. *Quarterly Journal of Economics* **117**, 817-869 (2002).
43  Fehr, E. & Schmidt, K. A theory of fairness, competition and cooperation. *Quarterly Journal of Economics* **114**, 817-868 (1999).
44  Bolton, G. E. & Ockenfels, A. ERC: A Theory of Equity, Reciprocity, and Competition. *The American Economic Review* **90**, 166-193 (2000).
45  Levine, D. K. Modeling Altruism and Spitefulness in Experiments. *Review of Economic Dynamics* **1**, 593-622 (1998).
46  Kahneman, D. A perspective on judgment and choice: Mapping bounded rationality. *American Psychologist* **58**, 697-720 (2003).
47  Kahneman, D. *Thinking, Fast and Slow*.  (Farrar, Straus and Giroux, 2011).
48  Gigerenzer, G. & Goldstein, D. G. Reasoning the fast and frugal way: models of bounded rationality. *Psychological review* **103**, 650 (1996).
49  Gigerenzer, G., Todd, P. M. & Group, A. R. *Simple heuristics that make us smart*.  (Oxford University Press, 1999).
50  Rand, D. G. *et al.* Social Heuristics Shape Intuitive Cooperation. *Nature Communications* **5**, Article number: 3677 (2014).
51  Peysakhovich, A., Nowak, M. A. & Rand, D. G. Humans Display a 'Cooperative Phenotype' that is Domain General and Temporally Stable. *Available at SSRN: http://ssrn.com/abstract=2426472* (2014).
52  Rand, D. G. & Kraft-Todd, G. T. Reflection Does Not Undermine Self-Interested Prosociality. *Frontiers in Behavioral Neuroscience* (In press).
53  Rand, D. G. & Gruber, J. Positive Emotion and (Dis)Inhibition Interact to Predict Cooperative Behavior. *Available at SSRN: http://ssrn.com/abstract=2429787* (2014).
54  Roberts, M. E. *et al.* Topic models for open ended survey responses with applications to experiments. *American Journal of Political Science* (In press).
55  Rand, D. G., Greene, J. D. & Nowak, M. A. Spontaneous giving and calculated greed. *Nature* **489**, 427-430 (2012).
56  Rand, D. G., Newman, G. E. & Wurzbacher, O. Social context and the dynamics of cooperative choice. *Journal of Behavioral Decision Making* (In press).
57  Rapoport, A. *Prisoner's dilemma: A study in conflict and cooperation*. Vol. 165 (University of Michigan Press, 1965).
58  Steele, M. W. & Tedeschi, J. T. Matrix indices and strategy choices in mixed-motive games. *Journal of Conflict Resolution*, 198-205 (1967).
59  Vlaev, I. & Chater, N. Game relativity: How context influences strategic decision making. *Journal of Experimental Psychology: Learning, Memory, and Cognition* **32**, 131 (2006).
60  Nowak, M. A. *Evolutionary dynamics: exploring the equations of life*. (Belknap press of Harvard University Press, 2006).
61  Blanco, M., Engelmann, D. & Normann, H. T. A within-subject analysis of other-regarding preferences. *Games and Economic Behavior* **72**, 321-338 (2011).
62  Paolacci, G., Chandler, J. & Ipeirotis, P. G. Running Experiments on Amazon Mechanical Turk. *Judgement and Decision Making* **5**, 411-419 (2010).
63  Rand, D. G. The promise of Mechanical Turk: How online labor markets can help theorists run behavioral experiments. *Journal of Theoretical Biology* **299**, 172-179 (2012).

64      Amir, O., Rand, D. G. & Gal, Y. K. Economic Games on the Internet: The Effect of $1 Stakes. *PLoS ONE* **7**, e31461 (2012).

**Author Contribution:** V.C., J.J.J., and D.G.R. designed the research, analysed the data and wrote the manuscript.

**Additional Information:** The authors declare no competing financial interests.

**Acknowledgements:** We would like to thank Zivvy Epstein and Grant Koplin for assistant with coding participant free-responses. Funding from the John Templeton Foundation through a subaward from the New Paths to Purpose Project at Chicago Booth is gratefully acknowledged.